\documentclass[%
 reprint,
 amsmath,amssymb,
 aps, prl,
]{revtex4-2}

\usepackage{morefloats}
\usepackage{color}
\usepackage{epsfig,graphicx,amsfonts,amsbsy}
\usepackage{amsmath,amsfonts,amsthm,amssymb}
\usepackage{appendix}
\usepackage{bbm}
\usepackage{makeidx}
\usepackage{url}
\usepackage{verbatim}
\usepackage{mathrsfs} 
\usepackage{morefloats}
\usepackage{comment}
\usepackage{appendix}
\usepackage{bbm}
\usepackage{makeidx}
\usepackage{url}
\usepackage{verbatim}
\usepackage[bookmarksnumbered,pdfpagelabels=true,plainpages=false,colorlinks=true,linkcolor=blue,citecolor=blue,urlcolor=blue]{hyperref}
\usepackage{array}
\usepackage{booktabs}
\usepackage{multirow}
\usepackage{bbm}
\usepackage{tabularx}
\usepackage{cancel,soul}
\usepackage{nicefrac, xfrac}
\usepackage{ulem}
\usepackage{bbold}
\usepackage{comment}

\usepackage[english]{babel}

\usepackage{amsmath}
\usepackage{graphicx}
\def \fcfm {Departamento de F\'isica, CEDENNA, FCFM, Universidad de Chile, Santiago, 8370448, Chile.}

\begin{document}
\title{Altermagnetic memcapacitors}
\author{Martin Latorre$^1$, and Alvaro S. Nunez$^1$}
\affiliation{$^1$\fcfm}

\begin{abstract}
We propose a spintronic memcapacitance effect based upon altermagnetic multiferroic materials. We identify the rare-earth vanadates RVO$_3$ as a concrete platform, with all key parameters tied to measured properties. Under an oscillating electric field, the resulting charge and spin
currents trace pinched hysteresis loops that close tangentially at
zero field -- the hallmark of memcapacitive, type-2 memdevice
behavior -- with charge current densities exceeding, by a factor of about
3.6, the lowest deterministic switching current density reported
for optimized spin-transfer-torque magnetic tunnel junctions. We model the system theoretically as a dimerized two-orbital $d$-wave altermagnetic lattice via a Su--Schrieffer--Heeger-type bond modulation, in the spirit of the spin-dependent Rice--Mele model, thereby coupling the altermagnetic order to field-switchable charge and spin polarizations. The associated polarization loops close tangentially at zero field and yield a sign-changing, history-dependent ``butterfly'' differential capacitance, identifying the device as a genuine memcapacitor. Both responses are protected by the same inversion symmetry, so charge and spin channels switch simultaneously with no separate control needed. These results establish altermagnetic multiferroics, realized concretely in RVO$_3$, as an efficient, non-volatile platform for combined electric and spintronic memory.
\end{abstract}

\maketitle
\section{Introduction}\label{sec1}

Multiferroic materials, in which two or more ferroic orders—most commonly
ferroelectricity, together with some form of magnetism, are present in a
single phase, has been a central topic in condensed matter physics over
the last twenty years \cite{Fiebig2005, SpaldinFiebig2005, Saez2023, Castro2024, Castro2025, Saez2024}. These systems
are compelling not only because of the complex underlying physics, where
lattice, charge, orbital, and spin degrees of freedom are strongly
interlinked, but also because of the technological potential of
magnetoelectric coupling: the ability to manipulate magnetic order using
an electric field, or conversely, to tune electric polarization with a
magnetic field, without incurring the dissipative losses associated with
passing currents through a material \cite{CheongMostovoy2007,Khomskii2009}.
The delineation of different multiferroic mechanisms—from improper,
spin-driven ferroelectricity to mechanisms involving lone pairs or
charge ordering—has led to a detailed classification scheme that still
shapes the exploration and design of new magnetoelectric materials.

In parallel, the emergence of altermagnetism has transformed the understanding of unconventional magnetic order. Altermagnets are collinear, fully compensated magnets with zero net magnetization, similar to conventional antiferromagnets, but their magnetic sublattices are related by rotational symmetries rather than by translational or inversion symmetries. This distinctive spin-group symmetry removes the spin degeneracy of electronic bands in momentum space, yielding $d$-, $g$-, or $i$-wave spin-split band structures even in the complete absence of spin-orbit coupling \cite{Smejkal2022a,Smejkal2022b}. Such nonrelativistic spin splitting imparts to altermagnets a range of phenomena previously regarded as unique to ferromagnets — including giant tunneling magnetoresistance, spin currents, and anomalous transport responses — while preserving the zero stray field and ultrafast spin dynamics that characterize antiferromagnets \cite{Smejkal2022c}. Recent symmetry analyses and spectroscopic studies have further clarified that this spin splitting does not have to arise from spin-orbit coupling; instead, it can be induced purely by orbital ordering in a Mott-insulating lattice \cite{jungwirth2026symmetry}, a mechanism that has been explicitly identified in correlated transition-metal oxides \cite{Leeb2024,Cuono2023}. Collectively, these features make altermagnets a compelling platform for spintronic technologies that unite high speed, scalability, and resilience to external magnetic fields \cite{JungwirthAltermagneticSpintronics,Zhang2026Staggered}.

The intersection of these two research directions—multiferroics and altermagnetism—has only recently begun to receive systematic attention. Because the altermagnetic order parameter is itself an anisotropic, ferroically ordered quantity (an even-parity, higher-order magnetic multipole, analogous to a magnetic octupole or a related tensorial object), it couples intrinsically to lattice strain and to polar structural distortions. Consequently, altermagnetism and ferroelectricity need not constitute independent ordering phenomena; instead, they can be cooperatively intertwined via the same underlying structural degrees of freedom \cite{BhowalSpaldin2024, smejkal2024altermagnetic}. 

Electric-field control of the associated spin splitting has already been demonstrated in a hybrid-improper molecular ferroelectric \cite{Gu2025_FSA}, and a Peierls-type bond dimerization has independently been proposed as the microscopic coupling mechanism in two-dimensional monolayer candidates \cite{Zhu2025_2DFEAM}. Collectively, these results establish that such a coupling is not merely a symmetry-mandated peculiarity, but rather a realistic and potentially robust pathway for electric-field control of spin transport. If this coupling can be realized or systematically engineered in actual materials, the resulting systems would combine electrically switchable altermagnetic spin splitting with the low-power, non-volatile control characteristic of ferroelectrics—a functionality that is unattainable in either conventional multiferroics or standard altermagnets considered in isolation.

Despite significant progress, a minimal, symmetry-transparent theoretical framework is still lacking to account for the \emph{simultaneous} emergence of altermagnetic, ferroelectric, and spin-transport (ferrospintronic) orders from a common microscopic mechanism, while being explicitly anchored to a concrete, experimentally accessible material platform. In this work, we address this deficiency by dimerizing the two-orbital altermagnetic lattice introduced in Ref.~\cite{Leeb2024} via a Su--Schrieffer--Heeger-type modulation of the hopping amplitudes, in direct analogy with the spin--Rice--Mele model \cite{Rice1982, Nunez2014, Ulloa2017, Vergara2024}. A single, chemically tunable dimerization parameter opens a spectral gap, breaks inversion symmetry, and thereby couples the pre-existing altermagnetic order to a finite electric polarization and spin polarization, both of which can be reversed by an applied electric field.

The model is deliberately constructed to be minimal: it isolates the essential symmetry constraints and microscopic ingredients required for the three orders to coexist and mutually interact. At the same time, it transcends the status of a purely academic toy model. We show that the same orbital-driven dimerization mechanism is realized and has been structurally characterized in the rare-earth vanadate series RVO$_3$, where the orbital-Peierls transition furnishes precisely the tunable structural control parameter demanded by theory. Within this unified framework, we demonstrate that these three orders are not merely compatible but dynamically coupled: an external electric field that reverses the ferroelectric polarization simultaneously reverses the associated spin polarization, both of which are tied to the same underlying, field-locked altermagnetic order. We further quantify this coupling by employing experimentally determined electronic and structural parameters specific to RVO$_3$.

This model quantitatively and microscopically establishes the efficiency of altermagnetic materials as memory elements. The same dimerization mechanism that couples electric and spin polarizations also generates a pinched, hysteretic current–field characteristic, such that a single structural degree of freedom—already present in an existing materials family—enables non-volatile, electrically switchable charge and spin memory without the need for additional device engineering. Consequently, altermagnetic multiferroics emerge not only as a new symmetry class to be classified, but as a design framework for discovering and optimizing practical, low-power memory elements at the interface of unconventional magnetism, ferroelectricity, and spintronics—an area that has thus far remained largely unexplored. 

Memcapacitors represent the second class in the family of memdevices (type-2 memdevices). They are memory-capacitive systems whose capacitance is not constant but varies with an internal state variable determined by the device’s past evolution. More precisely, an $n$-th order voltage-controlled memcapacitive system is described by
$q(t) = C(\mathbf{x}, V, t)\, V(t)$ together with $\dot{\mathbf{x}} =
f(\mathbf{x}, V, t)$, where $q$ denotes the charge, $V$ the applied voltage, $\mathbf{x}$ the set of $n$ internal state variables, and $C$ the memcapacitance \cite{Chua1971,Chua1976,Pershin01042011}. As for other memdevices, this state dependence gives rise to a pinched hysteresis loop in the charge–voltage characteristics under a time-dependent excitation
\cite{Chua1971,Chua1976,Pershin01042011,Lanza2025}. The complex dynamics and inherent memory properties of memdevices render them highly attractive components for contemporary computing paradigms, especially neuromorphic computing
\cite{Tetzlaff2013,Chua2019,DiVentra2023,Gaur2025,Liu2025,Song2024,Wang2018-gm}.

The remainder of this work is organized as follows. We first introduce the dimerized altermagnetic Hamiltonian and its coupling to an external electric field. We then derive the mean-field free energy and the driven dynamics of the associated order parameters, from which we compute the resulting charge and spin polarization currents. Finally, we identify RVO$_3$ as a concrete material platform that realizes this physics and characterize the corresponding altermagnetic memcapacitor response as a specific, experimentally relevant and energy-efficient memory unit.

\begin{figure*}[t]
    \centering
    \includegraphics[width=1\textwidth]{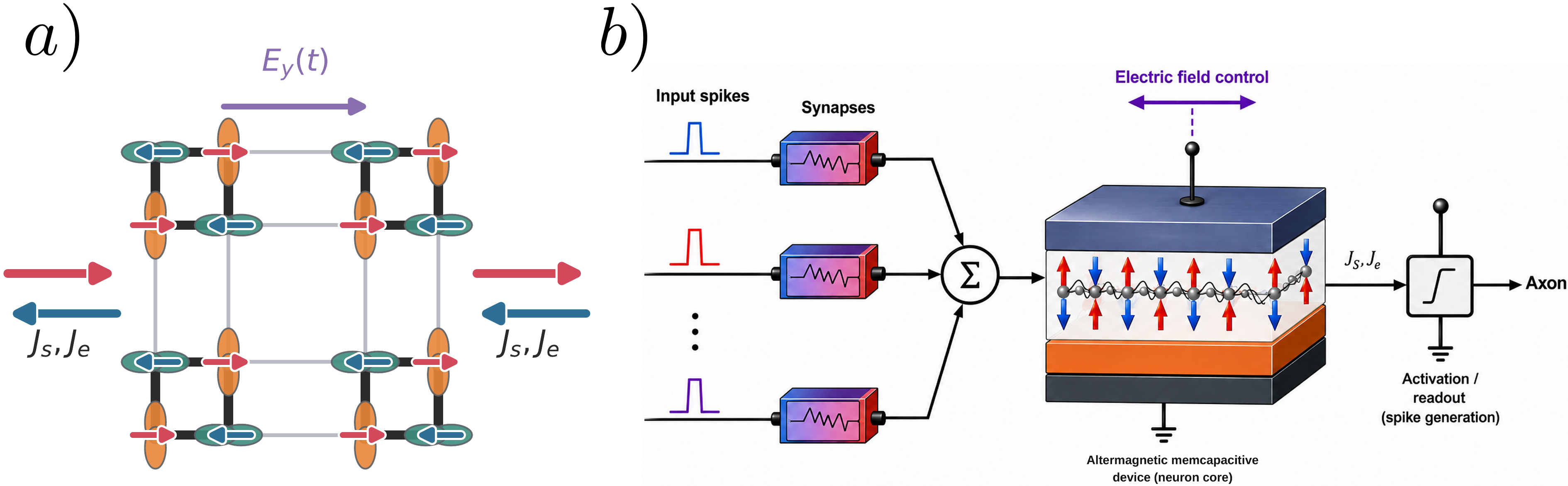}
    \caption{a) Pictographic representation of the coupled charge and
spin current emission under a time-dependent electric field
$E_y(t)$. Red and blue arrows denote spin up and down,
respectively; orange and green lobes denote the two altermagnetic
$d$-wave orbitals ($d_{xz}$, $d_{yz}$); bond thickness encodes the
SSH dimerization, with black (intra-cell) and grey (inter-cell)
bonds corresponding to the strong and weak hopping amplitudes
$t(1\pm\delta)$. Charge and spin currents $J_e, J_s$ are emitted
longitudinally, along the same direction as the driving field, on
the faces perpendicular to $E_y(t)$. b) Neuromorphic-neuron analogy
for the altermagnetic memcapacitive device: input spikes are
weighted by synaptic elements and summed ($\Sigma$) before driving
the altermagnetic memcapacitive device (neuron core), whose
field-controlled state $(m,o)$ sets the coupled charge--spin
response $J_e, J_s$; a downstream activation/readout stage converts
this response into an output spike (axon).}
    \label{fig:placeholder}
\end{figure*}

\section{Dimerized Altermagnetic Model.}
\label{sec:Dimerized Altermagnetic Model}
We base our analysis on the two-orbital square-lattice model introduced by Leeb \textit{et al.}~\cite{Leeb2024,DelaBarrera2025bias}. Despite its minimal character, this model reproduces the key  characteristics of $d$-wave altermagnets~\cite{ma}, namely a nonrelativistic, momentum-dependent spin splitting of the electronic bands arising from the concurrent presence of antiferromagnetic and orbital ordering. To incorporate polar degrees of freedom, we extend this framework to a dimerized altermagnet by implementing a Su--Schrieffer--Heeger (SSH) modulation of the nearest-neighbor hopping amplitudes. The resulting altermagnetic phase is analogous to a two-dimensional spin-dependent Rice--Mele Hamiltonian~\cite{Rice1982}, akin to those studied in Refs.~\cite{Nunez2014, Vergara2024, Saez2023, Saez2024, Castro2024}. The full Hamiltonian is given by
\begin{equation}
    \mathcal{H} = \mathcal{H}_t + \mathcal{H}_J
    + \mathcal{H}_V + \mathcal{H}_{\mathrm{pol}},
    \label{eq:Hfull}
\end{equation}
where $\psi^{\dagger}_{\mathbf{r}\alpha\sigma}$ denotes the creation operator for an electron at lattice site $\mathbf{r}=(i_x,i_y)$ with orbital index $\alpha \in \{d_{xz},d_{yz}\}$ and spin projection $\sigma \in \{\uparrow,\downarrow\}$.
We introduce dimerization into the kinetic sector by modulating the nearest-neighbor hopping amplitudes on $2\times2$ plaquettes according to $t_{1,2}\to t_{1,2}(1\pm\delta)$ on strong (intra-plaquette) and weak (inter-plaquette) bonds\cite{Vergara2024, Liu2017}. This modulation is governed by a single dimerization parameter $\delta$, while the diagonal hopping amplitudes $t_3$ and $t_4$—and consequently the altermagnetic spin splitting—are kept spatially uniform. Grouping the orbital degrees of freedom into a site spinor
$\Psi_{\mathbf{r}} = (\psi_{\mathbf{r}x\sigma},\psi_{\mathbf{r}y\sigma})^{\mathsf{T}}$,
for which the hopping is spin-diagonal, the kinetic Hamiltonian can be written as
\begin{equation}
    \mathcal{H}_t
    = -\sum_{\mathbf{r}}\sum_{\boldsymbol{\rho}}
    \bigl[\Psi^{\dagger}_{\mathbf{r}}\,\mathbf{t}_{\boldsymbol{\rho}}(\mathbf{r})\,
    \Psi_{\mathbf{r}+\boldsymbol{\rho}} + \mathrm{h.c.}\bigr],
    \label{eq:Htdim}
\end{equation}
where $\boldsymbol{\rho}\in\{\hat{x},\hat{y},\hat{x}+\hat{y},\hat{x}-\hat{y}\}$ denotes the four forward bond vectors to first- and second-neighbor sites, and the Hermitian conjugate generates the corresponding backward processes along $-\boldsymbol{\rho}$. The first-neighbor hopping matrices are orbital-diagonal, $\mathbf{t}_{\hat{x}} = (1\pm\delta)\,\mathrm{diag}(t_1,t_2)$ and $\mathbf{t}_{\hat{y}} = (1\pm\delta)\,\mathrm{diag}(t_2,t_1)$, thereby encoding both the orbital anisotropy and the dimerization. In contrast, the second-neighbor hopping matrices $\mathbf{t}_{\hat{x}\pm\hat{y}}$ are uniform and contain $t_3$ on the diagonal (intra-orbital hopping) and $\mp t_4$ on the off-diagonal (inter-orbital hopping), where the sign change between the two diagonal directions constitutes the characteristic $d$-wave structure of the altermagnet.

Choosing the $2\times2$ plaquette as the elementary unit cell leads to a folding of the Brillouin zone. In this representation, the dimerized kinetic Hamiltonian $\mathcal{H}_t^{\mathrm{dim}}$ is diagonal in the crystal momentum $\mathbf{K}$ of the reduced Brillouin zone and is described by a $16\times16$ Bloch matrix $h(\mathbf{K})$, corresponding to four sites, two orbitals, and two spin projections per cell. Strong (intra-plaquette) bonds connect sites within the same unit cell and thus yield $\mathbf{K}$-independent matrix elements, whereas weak (inter-plaquette) bonds connect adjacent unit cells and give rise to phase factors $e^{\pm iK_{x,y}}$. In the uniform limit $\delta\to0$, the Brillouin zone is unfolded and the original, non-dimerized band structure of Ref.~\cite{Leeb2024} is recovered.

The magnetic and orbital interactions are captured, respectively, by a
Heisenberg-like spin exchange and an Ising-like orbital exchange,
\begin{equation}
    \mathcal{H}_J = J\!\sum_{\langle\mathbf{r}\mathbf{r}'\rangle}
    \mathbf{S}_{\mathbf{r}}\!\cdot\!\mathbf{S}_{\mathbf{r}'},
    \quad
    \mathcal{H}_V = V\!\sum_{\langle\mathbf{r}\mathbf{r}'\rangle}
    N^{z}_{\mathbf{r}}N^{z}_{\mathbf{r}'},
    \label{eq:HJHV}
\end{equation}
where
$\mathbf{S}_{\mathbf{r}} = \sum_{\alpha\sigma\sigma'}\psi^{\dagger}_{\mathbf{r}\alpha\sigma}
\tfrac12\boldsymbol{\sigma}_{\sigma\sigma'}\psi_{\mathbf{r}\alpha\sigma'}$ is the local
spin operator and
$N^{z}_{\mathbf{r}} = \sum_{\alpha\beta\sigma}\psi^{\dagger}_{\mathbf{r}\alpha\sigma}
\tau^{z}_{\alpha\beta}\psi_{\mathbf{r}\beta\sigma}$ the orbital pseudospin operator, with
$\boldsymbol{\sigma}$ and $\tau^{z}$ the Pauli matrices acting on spin indices
$\sigma,\sigma'$ in $\{\uparrow,\downarrow\}$ and orbital indices
$\alpha,\beta$ in $\{d_{xz},d_{yz}\}$, respectively. In the altermagnetic region
$J  > 0$ drives $(\pi,\pi)$ antiferromagnetic order and
$V > 0$ drives $(\pi,\pi)$ orbital order; together these two
orders define the altermagnetic phase.

\section{Coupling to an external electric field.}
\label{sec:Coupling to an external electric field}
The final term in Eq.~\eqref{eq:Hfull} introduces the coupling to the driving field,
\begin{equation}
    \mathcal{H}_{\mathrm{pol}} = e\,a\,E_y\sum_{\mathbf{r}}\xi_{\mathbf{r}} \sum_{\alpha\sigma}\psi^{\dagger}_{\mathbf{r}\alpha\sigma}\psi_{\mathbf{r}\alpha\sigma},
    \label{eq:Hpol}
\end{equation}
with $\xi_{\mathbf{r}}=(-1)^{i_y}$ a $(0,\pi)$ sublattice modulation, odd under the plaquette-center inversion $\mathcal{P}$. This staggered on-site potential is generated by an external field $\mathbf{E}$ acting on the rigid lattice, obtained by projecting the scalar potential $\varphi(\mathbf{r})=-E_y\,y$ onto the $2\times2$ cell and keeping only the intracell staggered component $\propto \xi_{\mathbf r}$ that couples to the plaquette dipole $\mathbf p \parallel \hat y$. This field-induced potential is the two-dimensional generalization of the Rice--Mele mass, analogous to the topological multiferroic constructed in Ref.~\cite{Vergara2024}. Since the kinetic $\mathcal{H}_t^{\mathrm{dim}}$ and exchange $\mathcal{H}_J,\mathcal{H}_V$ terms are even under $\mathcal{P}$, only $\mathcal{H}_{\mathrm{pol}}$ breaks parity.

Once $m$ and $o$ are promoted to dynamical variables, $\mathcal H_{\mathrm{pol}}$ is the sole channel through which $E_y$ enters their equations of motion, as detailed below. This coupling is what generates the electrically switchable spin and charge polarization at the center of this work. We refer to the full $\mathcal{H}$ in Eq.~\eqref{eq:Hfull} as the dimerized altermagnetic model represented in Fig.~\ref{fig:placeholder}(a).

\section{Consistency with the Parent Altermagnetic Model}
\label{sec:consistency_check}

The phase diagram and spin-resolved band structure of
Sec.~\ref{sec:Dimerized Altermagnetic Model} were established in
Ref.~\cite{latorre2026electricallyswitchablespintronicsmultiferroic} for
representative values of the dimerization $\delta$ and field-induced
Rice--Mele mass $eaE_y$. Before building on that framework here, we verify
that the same qualitative behavior--gap opening and enlargement of the
altermagnetic region under simultaneous dimerization and field--is recovered
at the specific working point relevant to the present device architecture,
$\delta = 0.183$ and $eaE_y = 0.201\,t$ (with $t = 0.135$~eV), rather than at
the illustrative parameters used previously.

\begin{figure}[t]
    \centering
    \includegraphics[width=\columnwidth]{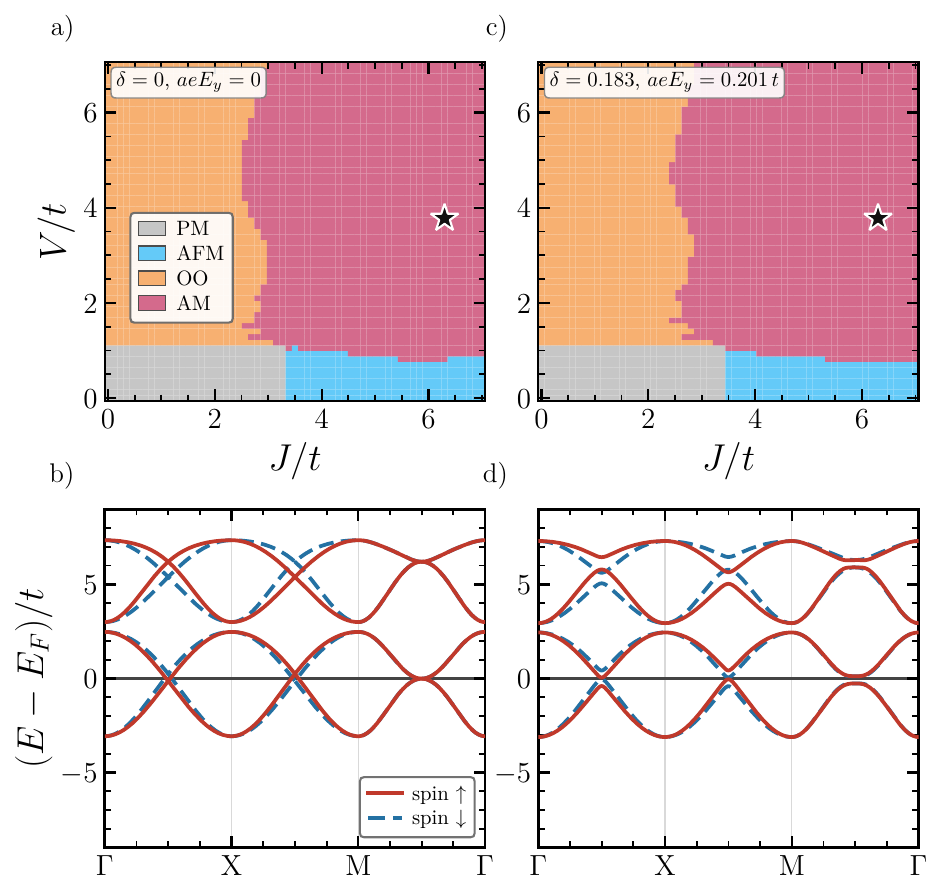}
    \caption{Consistency check of the dimerized altermagnetic model of
    Ref.~\cite{latorre2026electricallyswitchablespintronicsmultiferroic} at
    the working point used in this work. (a,c) Phase diagram in the $J$--$V$
    plane (units of $t$), with phases (PM, AFM, OO, AM) identified from the
    staggered amplitudes $(m,o)$; the star marks $J/t = 6.3$, $V/t = 3.78$.
    (b,d) Spin-resolved bands along $\Gamma$--X--M--$\Gamma$, as
    $(E-E_F)/t$, with spin~$\uparrow$ (solid red) and spin~$\downarrow$
    (dashed blue). Left column (a,b): undimerized, field-free reference,
    $\delta = 0$, $eaE_y = 0$ (indirect gap $-0.0513\,t$, metallic). Right
    column (c,d): working point of this work, $\delta = 0.183$,
    $eaE_y = 0.201\,t$ (indirect gap $+0.0844\,t$, direct gap $+0.3781\,t$,
    insulating). The altermagnetic fraction of the phase diagram increases
    from $52.3\%$ to $53.6\%$ between the two cases.}
    \label{fig:consistency_check}
\end{figure}

Repeating the mean-field calculation at fixed filling $1/8$ and
$J/t = 6.3$, $V/t = 3.78$ on a $61\times61$ grid, and comparing against the
undimerized, field-free case ($\delta = 0$, $E_y = 0$) on the same
$(J/t,V/t)$ grid [Fig.~\ref{fig:consistency_check}], we find the indirect
band gap changes sign, from $-0.0513\,t$ (overlapping bands, metallic) to
$+0.0844\,t$ (insulating), with a direct gap of $+0.3781\,t$
[Fig.~\ref{fig:consistency_check}(b,d)], while the staggered order
parameters remain essentially unchanged ($m$: $0.247\to0.246$; $o$:
$0.487\to0.486$). Over the same coupling range, the altermagnetic fraction
of the phase diagram [Fig.~\ref{fig:consistency_check}(a,c)] increases from
$52.3\%$ to $53.6\%$. Both effects reproduce, at this working point, the
mechanism reported in
Ref.~\cite{latorre2026electricallyswitchablespintronicsmultiferroic}:
dimerization and the polar field open a single-particle gap that removes
competing metallic screening, reinforcing rather than suppressing
altermagnetic order. This confirms that the model underlying the present
analysis behaves consistently at the parameters used throughout the rest of
this work.

\section{Free energy and driven dynamics.}
Within the mean-field decoupling of $\mathcal{H}_J$ and $\mathcal{H}_V$, the
staggered order parameters $m$ and $o$ are determined by imposing stationarity of a
Landau-like free energy functional
\begin{equation}
    F(m,o;E_y) = \frac{zJ}{2}\,m^2 + \frac{zV}{2}\,o^2
    + \Omega_{\mathrm{el}}(m,o;E_y),
    \label{eq:F}
\end{equation}
where $z=4$ denotes the lattice coordination number and
$\Omega_{\mathrm{el}}(m,o;E_y)$ is the grand-canonical potential of the
electronic subsystem at fixed $(m,o,E_y)$. The latter is obtained by
diagonalizing the Bloch Hamiltonian $h(\mathbf{K};m,o,E_y)$ in
Eq.~\eqref{eq:Htdim} and summing the occupied eigenvalues at fixed filling
and temperature. According to the Hellmann--Feynman theorem, the electronic
contribution incorporates the mean-field fields $B_s(m)=-zJm$ and
$B_o(o)=-zVo$ entering the on-site potential of $h$, such that demanding
stationarity of Eq.~\eqref{eq:F} leads to
\begin{equation}
\label{eq:saddle}
\begin{aligned}
    \frac{\partial F}{\partial m} &= zJ\bigl[m-\langle M\rangle(m,o,E_y)\bigr]=0, \\
    \frac{\partial F}{\partial o} &= zV\bigl[o-\langle O\rangle(m,o,E_y)\bigr]=0,
\end{aligned}
\end{equation}
is equivalent to the familiar self-consistency conditions $m=\langle
M\rangle$, $o=\langle O\rangle$, with $\langle M\rangle(m,o,E_y)$ and
$\langle O\rangle(m,o,E_y)$ the staggered spin and orbital expectation
values evaluated on the occupied Bloch states of $h(\mathbf{K};m,o,E_y)$ at
the trial point $(m,o,E_y)$, \emph{without} iterating to self-consistency.

Instead of constraining the system to follow this equilibrium manifold adiabatically, we allow $m$ and $o$ to relax towards it dynamically via the same phenomenological mechanism introduced in Ref.~\cite{Latorre2026pss}:
\begin{equation}
\begin{split}
    \tau_m\,\dot{m} &= -\frac{1}{zJ}\frac{\partial F}{\partial m}
    = \langle M\rangle(m,o,E_y) - m,
    \\
    \tau_o\,\dot{o} &= -\frac{1}{zV}\frac{\partial F}{\partial o}
    = \langle O\rangle(m,o,E_y) - o,
\end{split}
\label{eq:eom}
\end{equation}
where $\tau_m$ and $\tau_o$ denote the phenomenological relaxation times associated with the spin and orbital sectors, respectively, each parameter effectively characterizing the many-body relaxation of the underlying electronic degrees of freedom. The external field enters Eq.~\eqref{eq:eom} solely through $E_y$, in direct correspondence with Eq.~\eqref{eq:Hpol}: it does not appear as an explicit driving force but only as a parameter that shifts the instantaneous target values $\langle M\rangle$ and $\langle O\rangle$ towards which $m$ and $o$ relax. Under a harmonic drive $E_y(t) = E_0 \sin(\omega t)$, this coupled relaxation dynamics endows $m(t)$ and $o(t)$ with a memory of the field history: at any given time, they lag behind, rather than coincide with, the corresponding equilibrium values specified by Eq.~\eqref{eq:saddle}. Consequently, the trajectory traced by $(m,o)$ over a single drive period is generally non-retracing and can enclose a finite area in the $(m,o)$ plane.

\section{Response of the polarization.}
For a given choice of the order parameters $(m,o)$ at fixed external field $E_y$, the modern theory of polarization assigns to the occupied Bloch manifold of $h(\mathbf{K};m,o,E_y)$ a charge polarization $P_e$, defined as the Berry phase of the occupied bands. The spin polarization $P_s$ is obtained in complete analogy, by inserting the Pauli matrix $\sigma^z$ as a weight within the occupied manifold prior to performing the momentum integration. In this way, $P_e$ and $P_s$ probe, respectively, the total and the spin-resolved Berry phases associated with the same underlying band structure.

Since $m$ and $o$ become dynamical variables through Eq.~\eqref{eq:eom}, the polarizations $P_e(m,o)$ and $P_s(m,o)$ also acquire an explicit time dependence: as the order parameters evolve under the external drive, the polarization is concomitantly transported, and its time derivative defines a polarization current. Here $P_c$ is simply the standard Berry-phase polarization of the occupied Bloch manifold, computed following the procedure of Ref.~\cite{Liu2017}, but now evaluated on a manifold that depends parametrically on $(m,o)$ via the mean-field Hamiltonian. Because $P_c$ depends on time only through $m(t)$ and $o(t)$, the chain rule immediately yields
\begin{equation}
    I_c = \frac{\mathrm{d}P_c}{\mathrm{d}t}
    = \frac{\mathrm{d}m}{\mathrm{d}t}\,\frac{\partial P_c}{\partial m}
    + \frac{\mathrm{d}o}{\mathrm{d}t}\,\frac{\partial P_c}{\partial o},
    \qquad c=e,s,
    \label{eq:current}
\end{equation}
that is, the current decomposes into contributions from the two order-parameter channels, weighted by their respective rates of change and by the sensitivity of the Berry phase to each order parameter.
This is the same formal construction that produces the polarization current of the intrinsic piezoelectric memristor in Ref.~\cite{Latorre2026pss}, here generalized to the pair of altermagnetic order parameters $(m,o)$ and extended from a single charge channel to the combined charge and spin transport channels supported by this band structure.

This generalization has an immediate implication for the observed response. Under plaquette-center inversion $\mathcal{P}$, the electric field transforms as $E_y \to -E_y$, while the order parameters $m$ and $o$, both defined as bilinears that are even under $\mathcal{P}$, remain invariant; consequently the target densities $\langle M\rangle(m,o;E_y)$ and $\langle O\rangle(m,o;E_y)$ entering the relaxation dynamics of Eq.~\eqref{eq:eom} inherit this parity,
$
\langle M\rangle(m,o;-E_y) = \langle M\rangle(m,o;E_y)$ and
$
\langle O\rangle(m,o;-E_y) = \langle O\rangle(m,o;E_y).
$
In the driven cycle, the polarization and current loops of Fig.~4 are correspondingly pinched at $E_y=0$: as in any memdevice, the constitutive state variables $(m,o)$ retrace a common value at the zero crossing of the drive, so $P_c$ and $I_c$ coincide there for the forward and reverse branches, independently of the drive amplitude and frequency and of the relaxation times $\tau_m$ and $\tau_o$.

\section{Results}
\subsection{RVO$_3$ as an altermagnetic material platform}
\label{sec:material-rvo3}

Realizing the physics described above calls for a material family in
which the underlying lattice can be pushed continuously between a
non-dimerized and a dimerized state, while remaining, throughout, a
correlated $d$-orbital magnetic insulator. The rare-earth vanadates
RVO$_3$ (R = rare-earth ion or Y) offer exactly this kind of tunable
platform. Across this perovskite series, the vanadium $3d$ orbitals
support a rich orbital degree of freedom that couples strongly to the
lattice, and substitution of the rare-earth ion acts as a continuous,
chemically-controlled handle on the octahedral network: as the
rare-earth radius is varied, the corner-shared VO$_6$ octahedra rotate
and tilt, and the vanadium chains respond by developing a
bond-length alternation along their axis \cite{Ulrich2003,PhysRevLett.87.245501}.
This alternation is not a subtle effect -- it is directly resolved by
neutron and x-ray scattering as a genuine structural dimerization
accompanying an orbital-ordering transition \cite{Radhakrishnan2024},
and it emerges from an entropy-driven, orbital-Peierls mechanism rooted
in spin-orbital exchange physics that is by now well understood in this
class of Mott insulators \cite{Horsch2003,Sirker_2003,PhysRevB.75.184434}.

The appeal of this family is precisely that the degree of dimerization
is not fixed but graded across its members: moving along the
lanthanide series retunes the tolerance factor smoothly, taking the
family continuously from essentially undimerized, textbook
spin-orbital compounds \cite{Zhang2022} to strongly dimerized members
farther along the series, with the full magnetic and orbital phase
diagram mapped out both experimentally and from first principles
\cite{Miyasaka2003,Sasani2021}. This makes RVO$_3$ less a single
candidate compound than a continuously-tunable materials family in
which our order parameter can, in principle, be dialed by chemistry
alone -- turning a theoretical control knob into an experimentally
accessible axis, and offering a natural set of internal reference
compounds at the family's undimerized end.

Beyond hosting the structural dimerization our model relies on, this
same family of correlated, orbitally-active vanadates has recently
been identified as a genuine altermagnetic material platform: orbital
order of the kind ubiquitous in Mott-insulating transition-metal
oxides has been shown to generate altermagnetic spin splitting through
symmetry alone, without any need for spin-orbit coupling
\cite{Leeb2024}, and first-principles calculations
confirm this altermagnetic character directly in the vanadate lattice,
with the symmetry of the resulting spin-split bands set by the
underlying magnetic order \cite{Cuono2023}. The same orbital-driven
mechanism extends naturally to chemically related correlated oxides,
where it further generates large anomalous transport responses
\cite{PhysRevB.108.115138}, underscoring that Mott-insulating oxides
with active orbital degrees of freedom form a broad and versatile
family of altermagnetic materials rather than a collection of isolated
exceptions. RVO$_3$ therefore combines, within a single well-studied
materials family, a chemically tunable structural dimerization and an
intrinsically altermagnetic electronic structure -- the two ingredients
our model couples together -- making it a natural and experimentally
grounded testbed for the electrically switchable multiferroic
altermagnetism proposed here. We confirm below that the microscopic
model retains its gap-opening, altermagnetism-enhancing response when
evaluated directly at the coupling parameters representative of this
compound (Fig.~\ref{fig:consistency_check}).

\subsection{Altermagnetic memcapacitor}
\label{sec:memcapacitor}

\begin{figure*}[t]
    \centering
    \includegraphics[width=\linewidth]{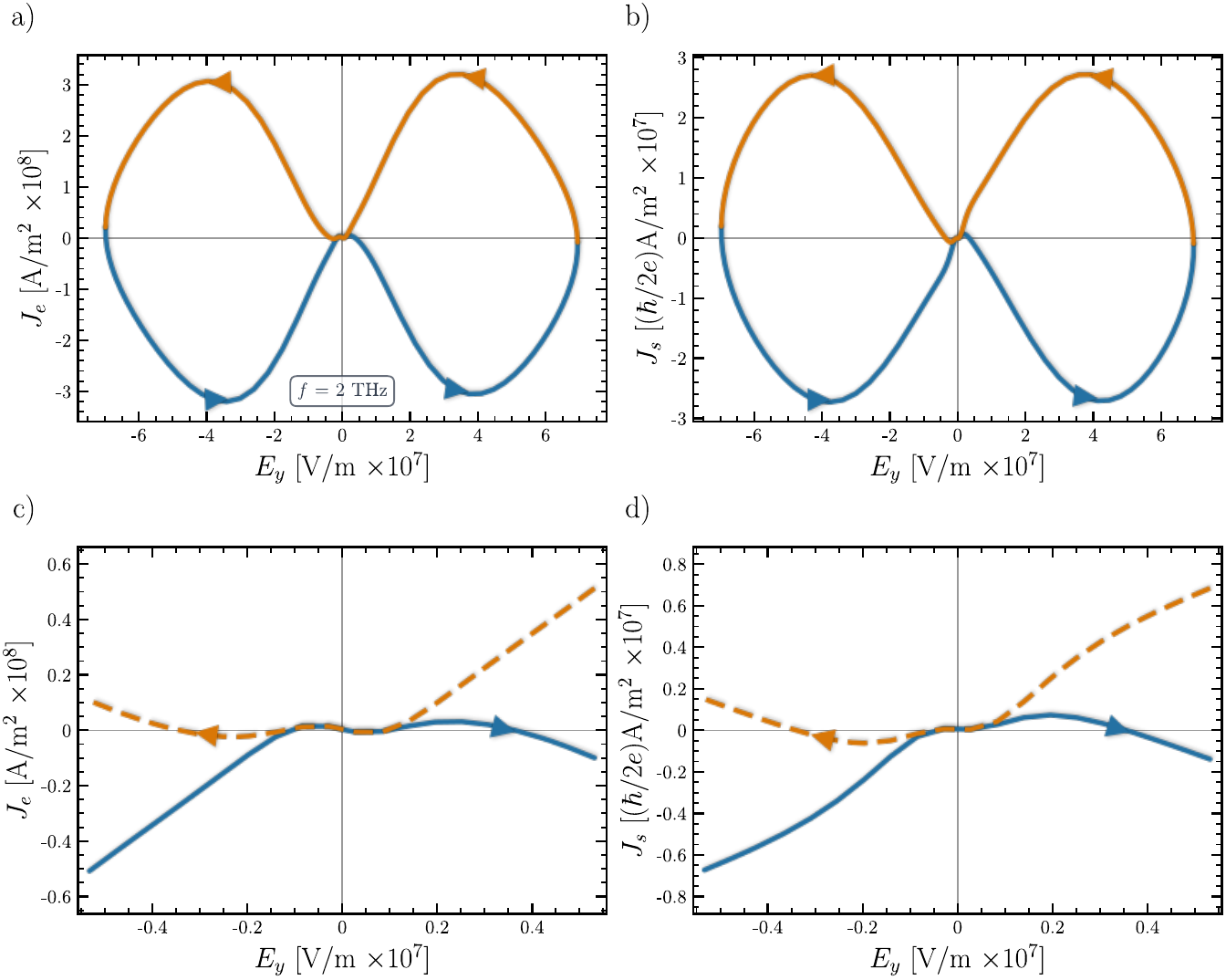}
    \caption{Pinched charge- and spin-current hysteresis loops of the
    altermagnetic memcapacitor, computed for the dimerized RVO$_3$ model at
    drive frequency $f=2$~THz. (a) Charge current density $J_e$ and (b)
    spin current density $J_s$ against the driving field
    $E_y(t)=E_0\sin(2\pi ft)$, $E_0=7\times10^{7}$~V/m; both loops are
    pinched at the origin by the inversion symmetry and reach
    $|J_e|_{\max}=3.2\times10^{8}$~A/m$^2$
    ($3.2\times10^{4}$~A/cm$^2$),
    $|J_s|_{\max}=2.7\times10^{7}\,(\hbar/2e)$~A/m$^2$, obtained from the
    sheet current of the two-dimensional tight-binding calculation via
    the interlayer spacing $c/2$ of the real RVO$_3$ stacking. Model
    parameters: dimerization $\delta=0.183$, spin exchange $J/t=6.3$,
    orbital exchange $V/t=3.78$, filling $n=1/8$ (one occupied band per
    spin block), working temperature $T/t=0.074$; material scale
    $t=0.135$~eV, lattice constant $a=3.88$~\AA, spin relaxation time
    $\tau_m=0.21$~ps, orbital relaxation time $\tau_o=0.004$~ps
    ($\rho=\tau_o/\tau_m=0.019$). (c)-(d) Close-up of panels (a) and (b), respectively, near
    $E_y=0$, showing the two branches merging tangentially — with a common
    slope, not a crossing corner.}
    \label{fig:memcapacitor}
\end{figure*}

Figure~\ref{fig:memcapacitor} shows the central result of this work: the
charge and spin current densities $J_e$ and $J_s$ generated by the
order-parameter channel of Eq.~\eqref{eq:current}, plotted against the
driving field $E_y(t)=E_0\sin(2\pi f t)$ over a full cycle at
$f=2$~THz. Both observables trace a pinched hysteresis loop that closes
tangentially at the origin, rather than a single-valued response
curve: as $E_y$ is swept from its negative to its positive extremum
and back, $J_e$ and $J_s$ follow different branches depending on
the sign of $\dot{E}_y$, so that a given field value $E_y$ is
compatible with two different current values -- the signature of a
memcapacitor, type-2 memdevice element, whose instantaneous
response depends on the recent history of the drive, not only on
its present value. The two branches merge tangentially -- with a
common slope, not a crossing corner -- at every zero crossing of
the drive and regardless of drive amplitude or frequency, as
required by the inversion symmetry discussed above; this tangential
pinch is what identifies the response as genuinely memcapacitive
rather than merely hysteretic. Panel
(a) shows the charge channel, with $J_e$ reaching
$\pm3.2\times10^{8}$~A/m$^2$ ($\pm3.2\times10^{4}$~A/cm$^2$) at the loop
extrema; panel (b) shows the spin channel, reaching
$\pm2.7\times10^{7}\,(\hbar/2e)$~A/m$^2$, over the same field sweep of
$E_y\in[-7,7]\times10^{7}$~V/m. That both channels are activated
simultaneously by the same field, with the same pinch protected by the
same symmetry, is a direct consequence of the shared origin of $J_e$ and
$J_s$ in the coupled relaxation of the altermagnetic and orbital order
parameters: the device switches charge and spin current in the same
stroke, with no separate control needed for either channel.

The values entering this calculation are not free fitting parameters but
are tied, wherever possible, to measured properties of the RVO$_3$
family. The dimerization $\delta=0.183$ and the exchange ratios
$J/t=6.3$, $V/t=3.78$ follow from the alternating superexchange constants
resolved along the vanadium chain \cite{Ulrich2003,Horsch2003}, and the
hopping scale $t=0.135$~eV follows from the same superexchange energy
combined with the on-site Coulomb repulsion of the $V^{3+}$ ion through
the standard superexchange relation, $J\sim4t^2/U$. The lattice constant
$a=3.88$~\AA\ and the working temperature, well below the family's
magnetic ordering temperature, are likewise structural and thermodynamic
properties of the compound rather than adjustable inputs. The only two
parameters that are not fixed by the electronic structure are the
relaxation times $\tau_m$ and $\tau_o$: because they describe the
dissipative coupling of the order parameters to the underlying electronic
bath, a coupling not captured by the mean-field electronic problem, they
are treated phenomenologically. Their ratio $\rho=\tau_o/\tau_m=0.019$
reflects the physically expected separation of time scales between
orbital and spin relaxation -- orbital relaxation, being lattice-mediated,
proceeds roughly two orders of magnitude faster than the spin relaxation
that ultimately paces the loop -- and it is this separation, not their
absolute values, that controls the shape and area of the hysteresis loop.

The values of $J_e$ and $J_s$ reported here are genuine three-dimensional
current densities, obtained from the sheet (per-unit-length) current
natively produced by the two-dimensional tight-binding calculation by
dividing through the interlayer spacing $c/2$ of the real RVO$_3$
stacking, under the decoupled-layers assumption stated above -- no
bundling of parallel chains is required to reach a density directly
comparable to three-dimensional switching experiments. At the loop
extrema, $J_e^{\max}=3.2\times10^{4}$~A/cm$^2$, a current density that
already exceeds, by a factor of about 3.6, the lowest deterministic
spin-transfer-torque switching current density reported for optimized
MgO-based perpendicular magnetic tunnel junctions, $9$~kA/cm$^2$
\cite{LeutenantsmeyerCurrents}. The intrinsic current response of the
platform is therefore, already at the single-chain level and without any
geometric rescaling, well within the regime required for practical
electrical switching of real spintronic devices.

\begin{figure*}[t]
    \centering
    \includegraphics[width=\linewidth]{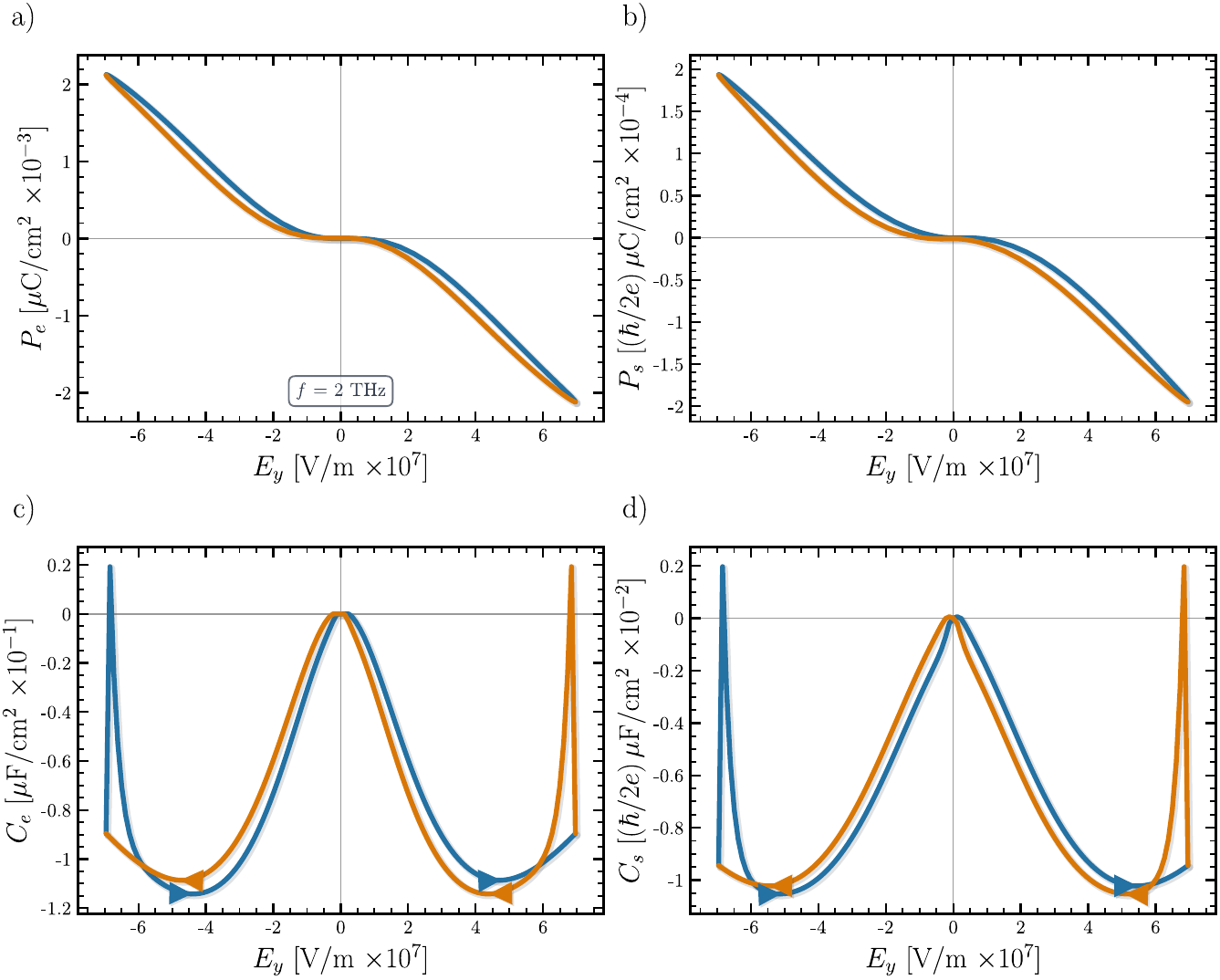}
    \caption{Polarization and differential-capacitance loops of the
altermagnetic memcapacitor at $f=2$~THz. (a) Charge-channel
polarization $P_e(m,o)$ and (b) spin-channel polarization
$P_s(m,o)$, both closing tangentially at the origin. (c) Charge-channel
differential capacitance $C_e = \mathrm{d}Q_e/\mathrm{d}E_y$ and (d) spin-channel
differential capacitance $C_s = \mathrm{d}Q_s/\mathrm{d}E_y$, obtained via the chain
rule through $(m,o)$ from the derivative of the loops in (a)-(b) and traced
against the driving field $E_y$; both vanish at $E_y=0$ and develop the
sign-changing, two-branch ``butterfly'' shape characteristic of a
voltage-controlled, history-dependent capacitance.}
    \label{fig:pol_cap}
\end{figure*}

\subsection{Capacitance behaviour}
\label{sec:capacitance_behaviour}

Figure~\ref{fig:pol_cap} shows the polarization loops $P_{e,s}(m,o)$
[panels (a)-(b)] and the differential-capacitance loops
$C_{e,s} = \mathrm{d}Q_{e,s}/\mathrm{d}E_y$ [panels (c)-(d)] obtained from the same
driven cycle at $f=2$~THz, both traced against the driving field $E_y$
although, as established above, the dependence on $E_y$ enters only
through the order parameters $(m,o)$. The polarization curves in
Fig.~\ref{fig:pol_cap}(a)-(b) display the characteristic S-shaped,
saturating response of a ferroelectric-like order parameter: $P_{e}$ and
$P_s$ grow steeply near $E_y=0$, develop an inflection point, and
saturate at $|P_e|_{\max}\approx2.1\times10^{-3}\,\mu\text{C/cm}^2$ and
$|P_s|_{\max}\approx1.95\times10^{-4}\,(\hbar/2e)\,\mu\text{C/cm}^2$ for
the largest driving fields. The forward and reverse branches
(blue/orange) are visibly split for intermediate $|E_y|$ and merge
\emph{tangentially} -- with a common slope, not a crossing corner -- at
$E_y=0$ and again at saturation, so the loop as a whole closes smoothly
through the origin rather than pinching it as a transversal crossing
would -- still the hallmark of a memory-circuit element rather than a
static, single-valued constitutive relation.

Differentiating this response with respect to the drive produces the
butterfly-shaped curves of Fig.~\ref{fig:pol_cap}(c)-(d). Both $C_e$
and $C_s$ are symmetric about $E_y=0$, where they vanish identically,
consistent with the pinching of the polarization and current loops at
$E_y=0$ discussed above; they then dip to a minimum,
$C_e\approx-1.15\times10^{-1}\,\mu\text{F/cm}^2$ and
$C_s\approx-1.03\times10^{-2}\,(\hbar/2e)\,\mu\text{F/cm}^2$, near the
coercive-like field $|E_y|\approx4\times10^{7}$~V/m -- where
$\mathrm{d}P/\mathrm{d}E_y$ is locally largest in magnitude -- before rising sharply back through zero
and turning positive as the drive approaches saturation. This
sign-changing, multi-valued $C$ vs. $E_y$ curve with two field-dependent branches is
the textbook capacitance-voltage ``butterfly'' signature of a
history-dependent, voltage-controlled capacitance: at a single value of
$E_y$ the differential capacitance takes two distinct values depending on
whether the system arrived from the positive or the negative saturation
branch, with both branches meeting tangentially rather than crossing at
$E_y=0$. The sharp upturn of $|C_{e,s}|$ near the turning points of the
drive is an artifact of the differential definition
$C_{\text{eff}}=I/(\mathrm{d}E_y/\mathrm{d}t)$ itself rather than a distinct physical
effect: because of the shape of the driving field, its rate $\mathrm{d}E_y/\mathrm{d}t$
vanishes at the extrema of the cycle while the order-channel current
$I_{e,s}$ remains finite there, so their ratio necessarily diverges at
those points. This divergence is therefore an inherent feature of how
the differential capacitance is defined near the reversal points of the
drive, not an additional signature of the memcapacitive response, and it
is expected to soften or shift with a different drive waveform or a
finite measurement bandwidth.

Within the classification of memcapacitive mechanisms introduced by
Pershin and Di Ventra \cite{Pershin01042011}, the present system
falls under the \emph{permittivity-related} class, and specifically
under the \emph{spontaneously-polarized medium mechanism}: the memory
effect does not arise from a time-changing plate geometry, nor from an
externally switched dielectric constant, but from an intrinsic,
spontaneously polarized order parameter -- here the altermagnetic
charge/spin polarization $P_{e,s}(m,o)$ -- whose state variable
$(m,o)$ evolves under the drive and thereby sets the instantaneous
capacitance. This is the same mechanism class historically associated
with ferroelectric capacitor structures, realized here through the
intrinsic order dynamics of the altermagnet rather than through an
auxiliary ferroelectric layer.

This behavior identifies the modeled order-channel response as a
\emph{memcapacitor} \cite{Pershin01042011}, in the sense formalized
by Di Ventra, Pershin and Chua: a capacitive element whose instantaneous
capacitance is a function of an internal state variable that encodes
the history of the drive, $C=C(x)$ with $\dot x$ set by the field-driven
order-parameter dynamics of Eq.~\eqref{eq:eom}. In our realization the
state variable is $x\equiv(m,o)$, so the dependence on $E_y$ enters
exclusively through $\dot x$ and not as a separate argument of $C$.
The two necessary fingerprints of such an element are both present
simultaneously in Fig.~\ref{fig:pol_cap}: (i) a charge-field loop that
closes tangentially at the origin, and (ii) a frequency-dependent,
non-single-valued $C$-$E$ curve collapsing toward that same tangent
point as $E_y\to0$. Physically, the sign change of $C_{e,s}$ across the
loop reflects a region of negative differential capacitance bracketing
the coercive field -- the same mechanism that underlies
capacitance-based memory windows in ferroelectric and multiferroic
capacitor stacks -- here obtained from the intrinsic altermagnetic order
dynamics rather than from an extrinsic dielectric layer.

\section{Conclusions}\label{sec5}

We have introduced a minimal, symmetry-transparent route to
electrically switchable multiferroic altermagnetism by dimerizing
a two-orbital $d$-wave altermagnetic lattice through a
Su--Schrieffer--Heeger-type bond modulation, in the spirit of the
spin-dependent Rice--Mele model. A single structural handle---the
dimerization $\delta$---simultaneously gaps the lattice, breaks
inversion symmetry, and locks the pre-existing altermagnetic order
to a finite, field-switchable charge and spin polarization. Unlike
purely symmetry-based constructions, every parameter entering the
model is tied to a measured property of the rare-earth vanadate
family RVO$_3$, whose orbital-Peierls structural dimerization and
intrinsically altermagnetic electronic structure supply, within a
single well-characterized materials platform, the two ingredients
our theory requires. Driven by an oscillating electric field, the
resulting charge and spin currents and their associated
polarizations close tangentially at zero field, the pinch required
of a genuine, type-2 memdevice rather than a merely hysteretic
response, and reach current densities that fall well within an
experimentally measurable range. Differentiating this response
yields a sign-changing, history-dependent ``butterfly''
differential capacitance, the second independent fingerprint that
identifies the device as a memcapacitor. Because both the current
and capacitance responses are protected by the same inversion
symmetry, the charge and spin channels switch simultaneously and
with a single control knob, without the need for separate
engineering of either degree of freedom.
 
These results reframe altermagnetic multiferroics, realized
concretely in RVO$_3$, as more than a new symmetry class to be
catalogued: they constitute a design principle for non-volatile,
electrically addressable memory elements that natively couple
charge and spin. This is precisely the functionality demanded by
emerging memcapacitive architectures for neuromorphic computing, and specifically by spin-based neuromorphic platforms that exploit spintronic memory elements as artificial synapses \cite{Marrows2024-iu,Kurenkov2020-kh},
where a device's instantaneous response must encode, rather than
merely record, the history of its input. As sketched in Fig. \ref{fig:placeholder}(b),
the altermagnetic memcapacitor is well suited to serve as the core
nonlinearity of an artificial neuron: input spikes summed at a
synaptic stage can drive the device's internal state $(m,o)$, whose
field-controlled relaxation continuously reshapes the coupled
charge--spin output $J_e, J_s$ that a downstream thresholding stage
converts into an output spike. Because the memcapacitance itself is
multi-valued and history-dependent, a single device is able to
combine synaptic weight storage and spin-resolved readout in one
element.
 
Several directions follow naturally from this work. Experimentally,
thin-film or heterostructure realizations of dimerized RVO$_3$
members would allow direct verification of the predicted pinched
current loops and butterfly capacitance under a gate or
out-of-plane field, and would clarify the role of strain,
substrate clamping, and finite-temperature fluctuations on the
coercive-like field and hysteresis area identified here.
Theoretically, extending the present mean-field, single-cell
treatment to include disorder, thermal noise, and device-to-device
variability will be essential for assessing the reliability of
memcapacitive synapses at array scale, as will a systematic study
of retention time and cycling endurance. More broadly, the same
orbital-driven dimerization mechanism is not unique to RVO$_3$: any
Mott-insulating altermagnet with an accessible orbital-Peierls or
analogous structural instability should support the same coupling,
opening a materials space considerably larger than the single
compound family explored here. Taken together, these results
establish altermagnetic multiferroics as an efficient,
materials-grounded platform for combined electric and spintronic
memory, and as a promising building block for low-power
neuromorphic hardware.

\paragraph{Acknowledgments} This work was funded by ANID CEDENNA CIA 250002.  Funding is acknowledged from Fondecyt Regular 1230515.
\bibliography{altermagnetic}

@article{Lanza2025,
  title = {The growing memristor industry},
  volume = {640},
  ISSN = {1476-4687},
  url = {http://dx.doi.org/10.1038/s41586-025-08733-5},
  DOI = {10.1038/s41586-025-08733-5},
  number = {8059},
  journal = {Nature},
  publisher = {Springer Science and Business Media LLC},
  author = {Lanza,  Mario and Pazos,  Sebastian and Aguirre,  Fernando and Sebastian,  Abu and Le Gallo,  Manuel and Alam,  Syed M. and Ikegawa,  Sumio and Yang,  J. Joshua and Vianello,  Elisa and Chang,  Meng-Fan and Molas,  Gabriel and Naveh,  Ishai and Ielmini,  Daniele and Liu,  Ming and Roldan,  Juan B.},
  year = {2025},
  month = apr,
  pages = {613–622}
}

@article{Smejkal2022a,
  author  = {\v{S}mejkal, Libor and Sinova, Jairo and Jungwirth, Tomas},
  title   = {Beyond Conventional Ferromagnetism and Antiferromagnetism: A Phase with Nonrelativistic Spin and Crystal Rotation Symmetry},
  journal = {Phys. Rev. X},
  volume  = {12},
  pages   = {031042},
  year    = {2022},
  doi     = {10.1103/PhysRevX.12.031042}
}

@article{Smejkal2022b,
  author  = {\v{S}mejkal, Libor and Sinova, Jairo and Jungwirth, Tomas},
  title   = {Emerging Research Landscape of Altermagnetism},
  journal = {Phys. Rev. X},
  volume  = {12},
  pages   = {040501},
  year    = {2022},
  doi     = {10.1103/PhysRevX.12.040501}
}

@article{Smejkal2022c,
  author  = {\v{S}mejkal, Libor and Hellenes, Anna Birk and Gonz\'alez-Hern\'andez, Rafael and Sinova, Jairo and Jungwirth, Tomas},
  title   = {Giant and Tunneling Magnetoresistance in Unconventional Collinear Antiferromagnets with Nonrelativistic Spin-Momentum Coupling},
  journal = {Phys. Rev. X},
  volume  = {12},
  pages   = {011028},
  year    = {2022},
  doi     = {10.1103/PhysRevX.12.011028}
}

@article{JungwirthAltermagneticSpintronics,
  author  = {Jungwirth, Tomas and Sinova, Jairo and Wadley, Peter and Kriegner, Dominik and Reichlov\'a, Helena and K\v{r}\'i\v{z}ek, Filip and Ohno, Hideo and \v{S}mejkal, Libor},
  title   = {Altermagnetic Spintronics},
  journal = {arXiv preprint arXiv:2508.09748},
  year    = {2025}
}

@article{Castro2024,
  title = {Toward Fully Multiferroic van der Waals SpinFETs: Basic Design and Quantum Calculations},
  volume = {24},
  ISSN = {1530-6992},
  url = {http://dx.doi.org/10.1021/acs.nanolett.4c01146},
  DOI = {10.1021/acs.nanolett.4c01146},
  number = {26},
  journal = {Nano Letters},
  publisher = {American Chemical Society (ACS)},
  author = {Castro,  Mario and Saéz,  Guidobeth and Vergara Apaz,  Patricio and Allende,  Sebastián and Nunez,  Alvaro S.},
  year = {2024},
  month = {June},
  pages = {7911–7918}
}

@article{Saez2023,
  title = {Model for Nonrelativistic Topological Multiferroic Matter},
  volume = {131},
  ISSN = {1079-7114},
  url = {http://dx.doi.org/10.1103/PhysRevLett.131.226801},
  DOI = {10.1103/physrevlett.131.226801},
  number = {22},
  journal = {Physical Review Letters},
  publisher = {American Physical Society (APS)},
  author = {Saez,  Guidobeth and Castro,  Mario A. and Allende,  Sebastian and Nunez,  Alvaro S.},
  year = {2023},
  month = {Nov}
}

@article{BhowalSpaldin2024,
  title = {Ferroically Ordered Magnetic Octupoles in $d$-Wave Altermagnets},
  author = {Bhowal, Sayantika and Spaldin, Nicola A.},
  journal = {Phys. Rev. X},
  volume = {14},
  issue = {1},
  pages = {011019},
  numpages = {19},
  year = {2024},
  month = {Feb},
  publisher = {American Physical Society},
  doi = {10.1103/PhysRevX.14.011019},
  url = {https://link.aps.org/doi/10.1103/PhysRevX.14.011019}
}

@article{Fiebig2005,
  author  = {Fiebig, Manfred},
  title   = {Revival of the Magnetoelectric Effect},
  journal = {J. Phys. D: Appl. Phys.},
  volume  = {38},
  pages   = {R123},
  year    = {2005},
  doi     = {10.1088/0022-3727/38/8/R01}
}

@article{SpaldinFiebig2005,
  author  = {Spaldin, Nicola A. and Fiebig, Manfred},
  title   = {The Renaissance of Magnetoelectric Multiferroics},
  journal = {Science},
  volume  = {309},
  pages   = {391--392},
  year    = {2005},
  doi     = {10.1126/science.1113357}
}

@article{CheongMostovoy2007,
  author  = {Cheong, Sang-Wook and Mostovoy, Maxim},
  title   = {Multiferroics: A Magnetic Twist for Ferroelectricity},
  journal = {Nat. Mater.},
  volume  = {6},
  pages   = {13--20},
  year    = {2007},
  doi     = {10.1038/nmat1804}
}

@article{Khomskii2009,
  author  = {Khomskii, Daniel},
  title   = {Classifying Multiferroics: Mechanisms and Effects},
  journal = {Physics},
  volume  = {2},
  pages   = {20},
  year    = {2009},
  doi     = {10.1103/Physics.2.20}
}

@article{Chua1971,
  author    = {Chua, Leon O.},
  title     = {Memristor---the missing circuit element},
  journal   = {IEEE Trans. Circuit Theory},
  volume    = {18},
  pages     = {507--519},
  year      = {1971},
  doi       = {10.1109/TCT.1971.1083337}
}

@article{Chua1976,
  author    = {Chua, Leon O. and Kang, Sung Mo},
  title     = {Memristive devices and systems},
  journal   = {Proc. IEEE},
  volume    = {64},
  pages     = {209--223},
  year      = {1976},
  doi       = {10.1109/PROC.1976.10092}
}

@article{Castro2025,
  title = {Phenomenological theory of electromagnons in multiferroic systems},
  volume = {111},
  ISSN = {2469-9969},
  url = {http://dx.doi.org/10.1103/PhysRevB.111.214401},
  DOI = {10.1103/physrevb.111.214401},
  number = {21},
  journal = {Physical Review B},
  publisher = {American Physical Society (APS)},
  author = {Castro,  Mario A. and Saji,  Carlos and Saez,  Guidobeth and Vergara,  Patricio and Allende,  Sebastian and Nunez,  Alvaro S.},
  year = {2025},
  month = {June} 
}

@article{Leeb2024,
  title = {Spontaneous Formation of Altermagnetism from Orbital Ordering},
  volume = {132},
  ISSN = {1079-7114},
  url = {http://dx.doi.org/10.1103/PhysRevLett.132.236701},
  DOI = {10.1103/physrevlett.132.236701},
  number = {23},
  journal = {Physical Review Letters},
  publisher = {American Physical Society (APS)},
  author = {Leeb,  Valentin and Mook,  Alexander and Šmejkal,  Libor and Knolle,  Johannes},
  year = {2024},
  month = {June} 
}

@article{DelaBarrera2025bias,
  title = {Electrical control of the exchange bias effect at model ferromagnet-altermagnet junctions},
  volume = {111},
  ISSN = {2469-9969},
  url = {http://dx.doi.org/10.1103/PhysRevB.111.174428},
  DOI = {10.1103/physrevb.111.174428},
  number = {17},
  journal = {Physical Review B},
  publisher = {American Physical Society (APS)},
  author = {De la Barrera,  Gaspar and Nunez,  Alvaro S.},
  year = {2025},
  month = may 
}

@article{Vergara2024,
  title   = {Emerging topological multiferroics from the two-dimensional Rice--Mele model},
  author  = {Vergara, Patricio and S{\'a}ez, Guidobeth and Castro, Mario and Allende, Sebasti{\'a}n and N{\'u}{\~n}ez, {\'A}lvaro S.},
  journal = {npj 2D Materials and Applications},
  volume  = {8},
  number  = {1},
  pages   = {41},
  year    = {2024},
  doi     = {10.1038/s41699-024-00478-5},
  publisher = {Nature Publishing Group}
}

@article{Liu2017,
  title   = {Novel Topological Phase with a Zero Berry Curvature},
  author  = {Liu, Feng and Wakabayashi, Katsunori},
  journal = {Phys. Rev. Lett.},
  volume  = {118},
  issue   = {7},
  pages   = {076803},
  year    = {2017},
  publisher = {American Physical Society},
  doi     = {10.1103/PhysRevLett.118.076803},
}

@article{Zhu2025_2DFEAM,
  title   = {Two-Dimensional Ferroelectric Altermagnets: From Model to Material Realization},
  author  = {Zhu, Ziye and Duan, Xunkai and Zhang, Jiayong and Hao, Bowen and {\v{Z}}uti{\'c}, Igor and Zhou, Tong},
  journal = {Nano Letters},
  volume  = {25},
  pages   = {9456--9462},
  year    = {2025},
  doi     = {10.1021/acs.nanolett.5c02121},
  publisher = {American Chemical Society},
}

@article{Gu2025_FSA,
  title   = {Ferroelectric Switchable Altermagnetism},
  author  = {Gu, Mingqiang and Liu, Yuntian and Zhu, Haiyuan and Yananose, Kunihiro and Chen, Xiaobing and Hu, Yongkang and Stroppa, Alessandro and Liu, Qihang},
  journal = {Phys. Rev. Lett.},
  volume  = {134},
  pages   = {106802},
  year    = {2025},
  doi     = {10.1103/PhysRevLett.134.106802},
  url     = {https://journals.aps.org/prl/abstract/10.1103/PhysRevLett.134.106802},
}

@misc{smejkal2024altermagnetic,
      title={Altermagnetic multiferroics and altermagnetoelectric effect}, 
      author={Libor Šmejkal},
      year={2024},
      eprint={2411.19928},
      archivePrefix={arXiv},
      primaryClass={cond-mat.mtrl-sci},
      url={https://arxiv.org/abs/2411.19928}, 
}

@BOOK{Tetzlaff2013,
  title     = "Memristors and Memristive Systems",
  editor    = "Tetzlaff, Ronald",
  publisher = "Springer",
  edition   =  2014,
  month     =  dec,
  year      =  2013,
  address   = "New York, NY",
  copyright = "https://www.springernature.com/gp/researchers/text-and-data-mining"
}

@BOOK{Chua2019,
  title     = "Handbook of memristor networks",
  editor    = "Chua, Leon and Sirakoulis, Georgios Ch and Adamatzky, Andrew",
  publisher = "Springer International Publishing",
  edition   =  2019,
  month     =  nov,
  year      =  2019,
  address   = "Cham, Switzerland"
}

@article{Rice1982,
  title = {Elementary Excitations of a Linearly Conjugated Diatomic Polymer},
  volume = {49},
  ISSN = {0031-9007},
  url = {http://dx.doi.org/10.1103/PhysRevLett.49.1455},
  DOI = {10.1103/physrevlett.49.1455},
  number = {19},
  journal = {Physical Review Letters},
  publisher = {American Physical Society (APS)},
  author = {Rice,  M. J. and Mele,  E. J.},
  year = {1982},
  month = nov,
  pages = {1455–1459}
}

@article{Nunez2014,
  title = {Theory of the piezo-spintronic effect},
  volume = {198},
  ISSN = {0038-1098},
  url = {http://dx.doi.org/10.1016/j.ssc.2013.10.018},
  DOI = {10.1016/j.ssc.2013.10.018},
  journal = {Solid State Communications},
  publisher = {Elsevier BV},
  author ={ {Nunez,  {Alvaro S.}}},
  year = {2014},
  month = nov,
  pages = {18–21}
}

@article{Ulloa2017,
  title = {Piezospintronic effect in honeycomb antiferromagnets},
  volume = {96},
  ISSN = {2469-9969},
  url = {http://dx.doi.org/10.1103/PhysRevB.96.104419},
  DOI = {10.1103/physrevb.96.104419},
  number = {10},
  journal = {Physical Review B},
  publisher = {American Physical Society (APS)},
  author = {Ulloa,  Camilo and Troncoso,  Roberto E. and Bender,  Scott A. and Duine,  R. A. and Nunez,  A. S.},
  year = {2017},
  month = {Sept} 
}

@article{Saez2024,
  title = {Ferrospintronic Order in Noncentrosymmetric Antiferromagnets: An Avenue toward Spintronic‐Based Computing,  Data Storage,  and Energy Harvesting},
  volume = {19},
  ISSN = {1862-6270},
  url = {http://dx.doi.org/10.1002/pssr.202400292},
  DOI = {10.1002/pssr.202400292},
  number = {3},
  journal = {physica status solidi (RRL) – Rapid Research Letters},
  publisher = {Wiley},
  author = {Saez,  Guidobeth and Vergara,  Patricio and Castro,  Mario and Allende,  Sebastian and Nunez,  Alvaro S.},
  year = {2024},
  month = nov 
}

@article{jungwirth2026symmetry,
  title={Symmetry, microscopy and spectroscopy signatures of altermagnetism},
  author={Jungwirth, Tomas and Sinova, Jairo and Fernandes, Rafael M. and Liu, Qihang and Watanabe, Hikaru and Murakami, Shuichi and Nakatsuji, Satoru and {\v{S}}mejkal, Libor},
  journal={Nature},
  volume={649},
  number={8098},
  pages={837--847},
  year={2026},
  publisher={Nature Publishing Group},
  doi={10.1038/s41586-025-09883-2},
  url={https://doi.org/10.1038/s41586-025-09883-2}
}

@article{ma,
  title = {Altermagnetic topological insulator and the selection rules},
  author = {Ma, Hai-Yang and Jia, Jin-Feng},
  journal = {Phys. Rev. B},
  volume = {110},
  issue = {6},
  pages = {064426},
  numpages = {6},
  year = {2024},
  month = {Aug},
  publisher = {American Physical Society},
  doi = {10.1103/PhysRevB.110.064426},
  url = {https://link.aps.org/doi/10.1103/PhysRevB.110.064426}
}

@article{Latorre2026pss,
  author  = {Latorre, Martin and De la Barrera, Gaspar and Troncoso, Roberto E. and Nunez, Alvaro S.},
  title   = {A Simple Model for an Intrinsic Piezoelectric Memristor},
  journal = {physica status solidi (RRL) -- Rapid Research Letters},
  volume  = {20},
  number  = {6},
  pages   = {e70197},
  year    = {2026},
  doi     = {10.1002/pssr.70197}
}

@article{Cuono2023,
  author  = {Cuono, Giuseppe and Sattigeri, Raghottam M. and Skolimowski, Jan and Autieri, Carmine},
  title   = {Orbital-Selective Altermagnetism and Correlation-Enhanced Spin-Splitting in Strongly-Correlated Transition Metal Oxides},
  journal = {Journal of Magnetism and Magnetic Materials},
  volume  = {586},
  pages   = {171163},
  year    = {2023},
  doi     = {10.1016/j.jmmm.2023.171163}
}

@article{Ulrich2003,
  author  = {Ulrich, C. and Khaliullin, G. and Sirker, J. and Reehuis, M. and Ohl, M. and Miyasaka, S. and Tokura, Y. and Keimer, B.},
  title   = {Magnetic Neutron Scattering Study of {YVO3}: Evidence for an Orbital Peierls State},
  journal = {Physical Review Letters},
  volume  = {91},
  pages   = {257202},
  year    = {2003},
  doi     = {10.1103/PhysRevLett.91.257202}
}

@article{Horsch2003,
  author  = {Horsch, P. and Khaliullin, G. and Ole\'s, A. M.},
  title   = {Dimerization versus Orbital-Moment Ordering in a Mott Insulator {YVO3}},
  journal = {Physical Review Letters},
  volume  = {91},
  pages   = {257203},
  year    = {2003},
  doi     = {10.1103/PhysRevLett.91.257203}
}

@article{Sirker_2003,
   title={Entropy driven dimerization in a one-dimensional spin-orbital model},
   volume={67},
   ISSN={1095-3795},
   url={http://dx.doi.org/10.1103/PhysRevB.67.100408},
   DOI={10.1103/physrevb.67.100408},
   number={10},
   journal={Physical Review B},
   publisher={American Physical Society (APS)},
   author={Sirker, J. and Khaliullin, G.},
   year={2003},
   month=Mar }

@article{PhysRevB.75.184434,
  title = {One-dimensional orbital fluctuations and the exotic magnetic properties of $\mathrm{Y}\mathrm{V}{\mathrm{O}}_{3}$},
  author = {Ole\ifmmode \acute{s}\else \'{s}\fi{}, Andrzej M. and Horsch, Peter and Khaliullin, Giniyat},
  journal = {Phys. Rev. B},
  volume = {75},
  issue = {18},
  pages = {184434},
  numpages = {21},
  year = {2007},
  month = {May},
  publisher = {American Physical Society},
  doi = {10.1103/PhysRevB.75.184434},
  url = {https://link.aps.org/doi/10.1103/PhysRevB.75.184434}
}

@article{PhysRevLett.87.245501,
  title = {Transition between Orbital Orderings in ${\mathrm{YVO}}_{3}$},
  author = {Blake, G. R. and Palstra, T. T. M. and Ren, Y. and Nugroho, A. A. and Menovsky, A. A.},
  journal = {Phys. Rev. Lett.},
  volume = {87},
  issue = {24},
  pages = {245501},
  numpages = {4},
  year = {2001},
  month = {Nov},
  publisher = {American Physical Society},
  doi = {10.1103/PhysRevLett.87.245501},
  url = {https://link.aps.org/doi/10.1103/PhysRevLett.87.245501}
}

@article{PhysRevB.108.115138,
  title = {Interplay between altermagnetism and nonsymmorphic symmetries generating large anomalous Hall conductivity by semi-Dirac points induced anticrossings},
  author = {Fakhredine, Amar and Sattigeri, Raghottam M. and Cuono, Giuseppe and Autieri, Carmine},
  journal = {Phys. Rev. B},
  volume = {108},
  issue = {11},
  pages = {115138},
  numpages = {8},
  year = {2023},
  month = {Sep},
  publisher = {American Physical Society},
  doi = {10.1103/PhysRevB.108.115138},
  url = {https://link.aps.org/doi/10.1103/PhysRevB.108.115138}
}

@article{Miyasaka2003,
  author  = {Miyasaka, S. and Okimoto, Y. and Iwama, M. and Tokura, Y.},
  title   = {Spin-Orbital Phase Diagram of Perovskite-Type {RVO3} (R = Rare-Earth Ion or Y)},
  journal = {Physical Review B},
  volume  = {68},
  pages   = {100406},
  year    = {2003},
  doi     = {10.1103/PhysRevB.68.100406}
}

@article{Zhang2022,
  author  = {Zhang, X.-J. and Koch, E. and Pavarini, E.},
  title   = {{LaVO3}: A True Kugel-Khomskii System},
  journal = {Physical Review B},
  volume  = {106},
  pages   = {115110},
  year    = {2022},
  doi     = {10.1103/PhysRevB.106.115110}
}

@article{Radhakrishnan2024,
  author  = {Radhakrishnan, P. and Rabinovich, K. S. and Boris, A. V. and F\"ursich, K. and Minola, M. and Christiani, G. and Logvenov, G. and Keimer, B. and Benckiser, E.},
  title   = {Imprinted Atomic Displacements Drive Spin--Orbital Order in a Vanadate Perovskite},
  journal = {Nature Physics},
  year    = {2024},
  doi     = {10.1038/s41567-024-02686-8}
}

@article{Sasani2021,
  author  = {Sasani, A. and \'I\~niguez, J. and Bousquet, E.},
  title   = {Magnetic Phase Diagram of Rare-Earth Orthorhombic Perovskite Oxides},
  journal = {Physical Review B},
  volume  = {104},
  pages   = {064431},
  year    = {2021},
  doi     = {10.1103/PhysRevB.104.064431}
}

@misc{leutenantsmeyerCurrents,
      title={Spin-transfer torque switching below 20 kA/cm$^2$ in perpendicular magnetic tunnel junctions}, 
      author={Johannes Christian Leutenantsmeyer and Marvin Walter and Steffen Wittrock and Patrick Peretzki and Henning Schuhmann and Michael Seibt and Markus Münzenberg},
      year={2013},
      eprint={1309.3397},
      archivePrefix={arXiv},
      primaryClass={cond-mat.mes-hall},
      url={https://arxiv.org/abs/1309.3397}, 
}

@article{Pershin01042011,
author = {Yuriy V. Pershin and Massimiliano Di Ventra},
title = {Memory effects in complex materials and nanoscale systems},
journal = {Advances in Physics},
volume = {60},
number = {2},
pages = {145--227},
year = {2011},
publisher = {Taylor \& Francis},
doi = {10.1080/00018732.2010.544961},


URL = { 
    
        https://doi.org/10.1080/00018732.2010.544961
    
    

},
eprint = { 
    
        https://doi.org/10.1080/00018732.2010.544961
    
    

}

}

@book{DiVentra2023,
  title = {Memristors and Memelements: Mathematics,  Physics and Fiction},
  ISBN = {9783031256257},
  ISSN = {2191-5431},
  url = {http://dx.doi.org/10.1007/978-3-031-25625-7},
  DOI = {10.1007/978-3-031-25625-7},
  journal = {SpringerBriefs in Physics},
  publisher = {Springer International Publishing},
  author = {Di Ventra,  Massimiliano and Pershin,  Yuriy V.},
  year = {2023}
}

@article{Gaur2025,
  title = {Molecularly Engineered Memristors for Reconfigurable Neuromorphic Functionalities},
  ISSN = {1521-4095},
  url = {http://dx.doi.org/10.1002/adma.202509143},
  DOI = {10.1002/adma.202509143},
  journal = {Advanced Materials},
  publisher = {Wiley},
  author = {Gaur,  Pallavi and Kundu,  Bidyabhusan and Ghosh,  Pradip and Bhattacharya,  Shayon and T,  Lohit and S,  Harivignesh and Rath,  Santi P. and Thompson,  Damien and Goswami,  Sreebrata and Goswami,  Sreetosh},
  year = {2025},
  month = dec 
}

@article{Song2024,
  title = {Dynamic Memristors for Temporal Signal Processing},
  volume = {9},
  ISSN = {2365-709X},
  url = {http://dx.doi.org/10.1002/admt.202400764},
  DOI = {10.1002/admt.202400764},
  number = {16},
  journal = {Advanced Materials Technologies},
  publisher = {Wiley},
  author = {Song,  Fuming and Shao,  He and Ming,  Jianyu and Sun,  Jintao and Li,  Wen and Yi,  Mingdong and Xie,  Linghai and Ling,  Haifeng},
  year = {2024},
  month = jul 
}

@article{Liu2025,
  title = {Memristors Based on 2D Materials: Bridging Device Theory and Potential Applications},
  volume = {36},
  ISSN = {1616-3028},
  url = {http://dx.doi.org/10.1002/adfm.202520816},
  DOI = {10.1002/adfm.202520816},
  number = {20},
  journal = {Advanced Functional Materials},
  publisher = {Wiley},
  author = {Liu,  Xueting and Pan,  Zhidong and Xia,  Zhonghui and Li,  Ling and Deng,  Qunrui and Pan,  Yuan and Li,  Jingbo and Huo,  Nengjie},
  year = {2025},
  month = oct 
}

@ARTICLE{Wang2018-gm,
  title     = "Bipolar analog memristors as artificial synapses for
               neuromorphic computing",
  author    = "Wang, Rui and Shi, Tuo and Zhang, Xumeng and Wang, Wei and Wei,
               Jinsong and Lu, Jian and Zhao, Xiaolong and Wu, Zuheng and Cao,
               Rongrong and Long, Shibing and Liu, Qi and Liu, Ming",
  journal   = "Materials (Basel)",
  publisher = "MDPI AG",
  volume    =  11,
  number    =  11,
  pages     = "E2102",
  month     =  oct,
  year      =  2018,
  copyright = "https://creativecommons.org/licenses/by/4.0/"
}

@ARTICLE{Marrows2024-iu,
  title     = "Neuromorphic computing with spintronics",
  author    = "Marrows, Christopher H and Barker, Joseph and Moore, Thomas A
               and Moorsom, Timothy",
  journal   = "Npj Spintron.",
  publisher = "Springer Science and Business Media LLC",
  volume    =  2,
  number    =  1,
  month     =  apr,
  year      =  2024,
  copyright = "https://creativecommons.org/licenses/by/4.0"
}

@ARTICLE{Kurenkov2020-kh,
  title     = "Neuromorphic computing with antiferromagnetic spintronics",
  author    = "Kurenkov, Aleksandr and Fukami, Shunsuke and Ohno, Hideo",
  journal   = "J. Appl. Phys.",
  publisher = "AIP Publishing",
  volume    =  128,
  number    =  1,
  pages     = "010902",
  month     =  jul,
  year      =  2020
}

@article{latorre2026electricallyswitchablespintronicsmultiferroic,
  author   = {Latorre, Martin and Allende, Sebastian and Nunez, Alvaro S.},
  title    = {Electrically Switchable Spintronics in a Multiferroic Altermagnet},
  journal = {arXiv preprint},
  eprint   = {2608.15953},
  archivePrefix = {arXiv},
  primaryClass  = {cond-mat.mes-hall},
  doi      = {10.48550/arXiv.2608.15953},
  year     = {2026},
  url      = {https://arxiv.org/abs/2608.15953}
}

@article{Zhang2026Staggered,
  author  = {Zhang, Jie and Fang, Ruijing and Zhou, Zhichao and Li, Xiao},
  title   = {Staggered Nonlinear Spin Generations in Centrosymmetric Altermagnets under Electric Current},
  journal = {Chinese Physics Letters},
  volume  = {43},
  number  = {2},
  pages   = {020711},
  year    = {2026},
  doi     = {10.1088/0256-307X/43/2/020711}
}

\end{document}